# Encoding complex fields by using a phase-only optical element: mitigation of pixel crosstalk effects

MIGUEL CARBONELL-LEAL,[1] AND OMEL MENDOZA-YERO,[1,*]

[1] GROC•UJI, Institute of New Imaging Technologies, Universitat Jaume I, 12071-Castelló, Spain
*Corresponding author: omendoza@fca.uji.es

**In this letter we report on the effects of pixel crosstalk on the experimental realization of a reported encoding method (Opt. Lett. 39, 1740 (2014)) with PA-LCoS SLMs. We found that, under Nyquist limit condition, about 70% of a single pixel cell can generate unexpected phase modulation. In order to approach uniform phase modulation, and consequently improve the quality of measured amplitude and phase images, a generalized sampling scheme is proposed. To corroborate our proposal, proper experiments were carried out. On this point, a particular implementation of the well-established phase shifting technique allows us to measure the retrieved complex field by using just a single camera.**

At present, there are a wide variety of reported optical methods addressed to optically manipulate the complex field of laser beams by using spatial light modulators [1-14]. Among them, those ones based on the use of parallel aligned liquid crystal on silicon (PA-LCoS) SLMs have gain particular attention not only because of their relative high efficiency, but also due to their proved ability to accurately modify the physical behavior of laser beams by using just a single phase element encoded into a phase-only SLM [3, 4, 7, 8-14]. Here, we deal with a particular interferometric method aimed to encode and retrieve the complex field of coherent laser beams that was introduced some few years ago [15]. This method has been widely and successfully employed in several experimental tasks including, but not limited to, demonstrate the Talbot self-imaging in the azimuthal angle [16], generate speckleless holographic displays [17], trap magnetic microparticles employing Bessel-Gauss beams [18], shape optical vector beams [19], or experimentally investigate the propagation and focusing characteristics of Airy beams [20, 21]. In all these applications the implementation of the above-mentioned encoding method [15] was carried out with the help of a commercially available PA-LCoS SLM. So, light undergoes phase modulation due to a change in the refractive index of the nematic liquid crystal (LC) material. Specifically, the phase shift is associated with the tilt of the SLM molecules when a signal voltage is applied between the front and the back faces of each LC cell. In addition, main design features of these devices ensure, in principle, that the phase modulation process is done with almost no coupling of amplitude modulation or change in the polarization state of the incident light, which is highly desirable for applications involving interference or diffraction phenomena.

However, PA-LCoS SLMs are not ideal devices, but show some unwanted effects that cause degradation of the phase modulation response. For instance, temporal fluctuations of the LC molecular orientation as a function of time causes depolarization effects, deteriorating the diffraction efficiency of SLM [22]. Another harmful effect is related to the Fabry-Perot multiple beam interference generated by the intrinsic layer structure of the LC device [23], that may originate non-linear phase modulation or even some coupling of amplitude modulation. In this context, there is a particular unwanted effect that becomes critical for applications that require encoded patterns with abrupt phase discontinuities, e.g., phase gratings with few-pixels period or high spatial frequency phase distributions associated with high scattering media. This effect can produce variations in the orientation of LC molecules at adjacent cells, and consequently modify its expected phase response. In the literature, this widely-studied phenomenon [24-27] is known as fringing field effect (or pixel crosstalk effect). In this context, as in the proposed method [15] the encoded phase element is computer generated by spatially multiplexing two different phase patterns at Nyquist limit, one should expect that its experimental realization with PA-LCoS SLMs does not lack from pixel crosstalk effects.

In this Letter we experimentally show that pixel crosstalk effects do deteriorate the amplitude and phase patterns obtained due to the application of the encoding method [15]. To alleviate these effects, a generalized sampling scheme able to significantly reduce non-uniform phase modulation without compromising the accuracy of the retrieved spectrum at the Fourier plane is proposed. In order to avoid the influence of the zero order coming from the SLM on the measurements done throughout this work, a fixed blazed grating encoded from $-\pi$ to $\pi$ was added to each phase element sent to the SLM, in the same manner as proposed in [17]. In addition, we also show how a polarization-based phase shifting technique [28] can be employed to measure both the amplitude and phase of the generated complex field using only a conventional camera.

The theory underlying the encoding method [15] can be briefly described as follow. Any complex field represented in the form $U(x,y) = A(x,y)e^{i\varphi(x,y)}$ can be also rewritten as:

$$U(x,y) = e^{i\theta(x,y)} + e^{i\vartheta(x,y)} \qquad (1)$$

where,

$$\theta(x,y) = \varphi(x,y) + \cos^{-1}[A(x,y)/A_{\max}] \qquad (2)$$

$$\vartheta(x,y) = \varphi(x,y) - \cos^{-1}[A(x,y)/A_{\max}] \qquad (3)$$

In Eqs. 1-3, the amplitude and phase of the two-dimensional complex field $U(x,y)$ is given by $A(x,y)$ and $\varphi(x,y)$, respectively.

In addition, $A_{max} \equiv 2$ holds for the maximum of $A(x,y)$. From Eq. (1), it is apparent that $U(x,y)$ can be obtained, from the coherent sum of uniform waves $e^{i\theta(x,y)}$ and $e^{i\vartheta(x,y)}$. To do that by using just a phase-only SLM, above uniform waves are spatially multiplexed (at Nyquist limit) with two-dimensional binary gratings in order to get a single phase element $\alpha(x,y)$ as follows:

$$M_1(x,y)e^{i\theta(x,y)} + M_2(x,y)e^{i\vartheta(x,y)} = e^{i\alpha(x,y)} \quad (4)$$

where,

$$\alpha(x,y) \equiv M_1(x,y)\theta(x,y) + M_2(x,y)\vartheta(x,y) \quad (5)$$

At this point, it should be noted that the interference of above-mentioned uniform waves cannot happen if we do not mix the phase information contained in $\alpha(x,y)$. This is carried out by using a spatial filter able to block all diffraction orders but the zero one. It can be shown that, after the filtering process, the spectrum of the original complex field can be exactly retrieved at the Fourier plane. Consequently, at the output plane of the imaging system, the retrieved complex field $U_{RET}(x,y)$, (without considering constant factors), is given by the convolution of the magnified and spatially reversed complex field $U(x,y)$ with the Fourier transform of the filter mask, that is:

$$U_{RET}(x,y) = U(-x/Mag, -y/Mag) \otimes F\{P(u,v)\} \quad (6)$$

In Eq. (6), the convolution operation is denoted by the symbol $\otimes$, and the term $Mag$ represents the magnification of the imaging system. So, from Eq. (6), in theory, the amplitude and phase of the original complex field is fully retrieved, except for some loss of spatial resolution due to the convolution operation.

On the other hand, the real physical behavior of phase-only SLM devices, under extreme pixel-to-pixel phase modulation conditions, can originate discrepancies between the theory and experiment. This mainly happens because the phase information associated with each uniform wave is spatially multiplexed at the Nyquist limit (pixel-cell 1). However, if the period of the binary gratings are not taken at the Nyquist limit, but the spatial frequency separations of diffraction orders are great enough to avoid overlapping among them, the Whittaker-Shannon sampling theorem ensures that, for bandlimited functions, the reconstruction of the spectrum at the Fourier plane is still accomplished exactly. In practice, the utilization of phase elements $\alpha(x,y)$ that are computer generated from binary gratings with pixel-cells greater than 1 should alleviate crosstalk effects.

In order to corroborate the above idea, we carry out a simply task. That is; we use our method to encode and measure a set of $N$ flat amplitude patterns under different pixel-cell configurations. In Fig. 1, the procedure used to digitally construct each phase element $\alpha(x,y)$ is schematically shown by means of three examples. The experimental data allows us to analyze the dependence of the pixel-cell size on the measured irradiance. In this scenario, Eq. (2) and (3) are reduced to the expressions: $\theta_s = \beta_s = \cos^{-1}(A_s/2)$, and $\vartheta_s = -\beta_s$, where $A_s$ with $s = 1,2,3...N$ holds for each amplitude pattern having an intensity or gray level that ranges from 0 to 2. The sum of uniform waves given by Eq. (1) is also simplified as:

$$e^{i\beta_s(x,y)} + e^{-i\beta_s(x,y)} = 2\cos(\beta_s) \quad (7)$$

In practice, we computer generate $N$ different phase elements $\alpha(x,y)$ from $N$ values of $\beta_s$ that ranges from 0 to $\pi$, and record corresponding images with a camera. The same procedure is repeated for different pixel-cell configurations. For each of them, the iris size was adjusted to block all frequencies but the zero one.

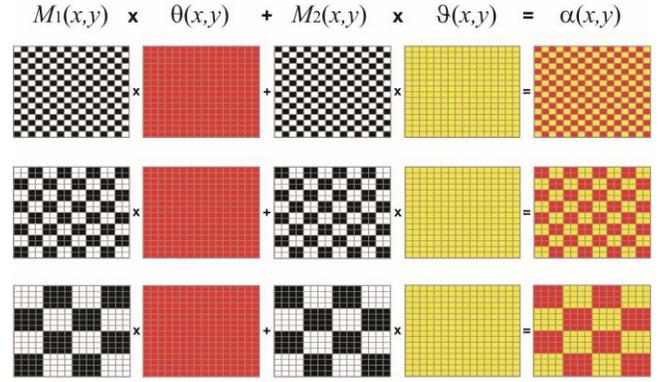

**Fig. 1.** (Color online) Example of the construction process of the phase element $\alpha(x,y)$ for different pixel-cell configurations.

The optical setup is shown in Fig. 2. We use a quasi-monochromatic laser beam of 10 nm spectral width and centered at 800 nm as light source. Before it impinges onto a reflective phase-only PA-LCoS SLM (Holoeye Pluto optimized for 700 nm-1000 nm, resolution 1920 × 1080 pixels, pixel pitch 8 μm, and phase range $3\pi$), the beam is conveniently attenuated with neutral filters (NF), and spatially magnified by using a commercial (BE06R) reflective 6X beam expander (BE). Then, the beam is sent to the SLM forming a small angle with the normal to the LC surface of about 4 degrees, and is back reflected towards the entrance of a 4f imaging system. The input plane of this optical system coincides with the SLM plane. The 4f imaging system is made up of a couple of refractive lenses (L1 and L2) with focal lengths of 1 m and 0.5 m, respectively. This pair of lenses gives a transversal demagnification of ½ at the output plane of the imaging system. This reduction allows us to directly measure the irradiances with a CMOS camera (Ueye UI-1540M, 1280 × 1024 pixel resolution and 5.2 pixel pitch). At the Fourier plane, the beam is transmitted through a low-pass spatial filter that consists of a hard circular iris (ID12Z/M).

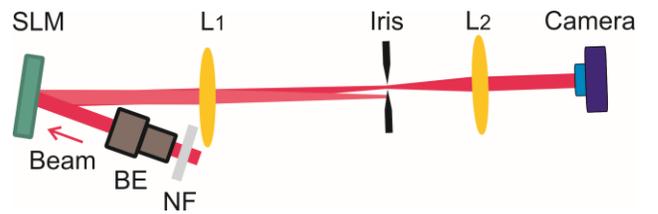

**Fig. 2.** (Color online) Off-axis optical configuration used to study pixel crosstalk effects in the proposed encoding method [15].

The experimental normalized irradiance curve, obtained under the Nyquist limit condition is shown in Fig. 3 with cross points. From the theoretical point of view, the corresponding irradiance $4[\cos(\beta_s)]^2$ given by Eq. (7) is represented in Fig. 3 with the thinnest curve (mentioned as ideal case in the inset box). After a visual inspection of both curves, it is apparent that experimental results do not reproduce the irradiance behavior predicted by the theory. Such a result clearly confirm that, for abrupt discontinuities in the encoded phase $\alpha(x,y)$, the SLM generates phase responses that not follows the grey-level/phase conversion derived from the manufacturer lookup table.

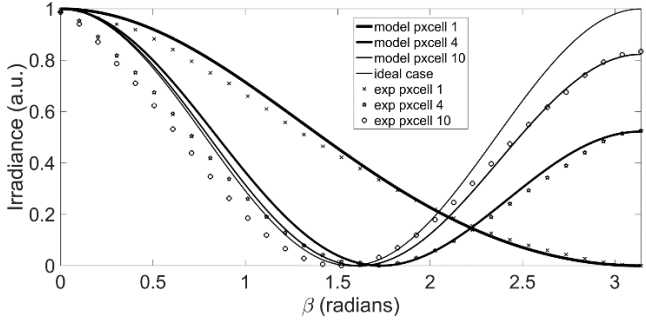

**Fig. 3.** Irradiance data points derived from the experimental measurements under pixel-cells 1, 4 and 10, and corresponding theoretical curves due to the proposed model. The ideal case is given by the expression $4[\cos(\beta_s)]^2$.

On the other hand, out of the Nyquist limit condition, as far as we increase the period of the sampling gratings, the measured irradiance curves (represented with star and circle points) begin to approach the theoretical one. This is shown in Fig. 3 with the two remaining experimental curves obtained for sampling gratings composed of 16 and 100 pixels per cell (or 4-, and 10- pixel-cell). These curves show a visible second maximum, but their peak values still decrease as the number of pixel per cell also do.

At this point, we theoretically verify that, in all cases, the behavior of measured reported in Fig. 3 can be mainly attributed to pixel crosstalk effects. To do that a simple modification in Eq. (7) was introduced. We assume that each pixel-cell (composed of one or more pixels) of the gratings is characterized by a non-uniform phase response. That is; each square pixel-cell can be thought to be divided into two different zones, the central zone composed of a concentric square of reduced area, and the remaining one that represents its border. At the central zone, the phase response of the pixel-cell is equal to that derived from the manufacturer lookup table, but at the border zone is reduced by half. In this way, the phase response at the central part of each pixel-cell does not change, and non-uniform phase modulation is accomplished only at the border of it. For the simulations, the ratio $\eta$ of areas corresponding to the border and central zones of each pixel-cell was used as a variable to adjust the measured data. In Fig. 3, we show (with continues lines of different thickness) the simulated irradiance curves obtained for each sampling grating. After a visual comparison of theoretical and experimental irradiance curves, it is apparent that our physical model agrees quite well with the experiment. In fact, the calculated root mean square error (rmse) between theory and experiment is less than 5% in all cases. Of course, there are small observable discrepancies between measured and predicted irradiance values (mostly in the first half part determined by the angle range) that our simple physical model cannot explain. Even so, we believe that it is enough to corroborate that observable facts are mostly caused by well-known [24-27] pixel crosstalk effects.

The simulated irradiance curves shown in Fig. 3 were generated with the ratios $\eta_{10}=0.09$, $\eta_4=0.22$, and $\eta_1=0.73$, that corresponds to sampling gratings of 10-, 4-, and 1- pixel-cells, respectively. Hence, in the worst-case scenario (Nyquist limit condition) we found that about 70% of each pixel does encode the phase information as manufacturer lookup table specifies. In contrast, if we increase a little bit the pixel-cell size, the phase response of our SLM can significantly approximates to the ideal behavior.

Now, in order to show how pixel crosstalk effects can influence the quality of the amplitude and phase images retrieved with the proposed encoding method [15], we carry out a couple of experimental tests. In the first one, an amplitude-only pattern given by a mushroom image is encoded by using the same sampling gratings as before. Consequently, three different phase elements $\alpha(x,y)$, corresponding to binary gratings with pixel-cells of 1, 4, and 10 were computer generated. Again, the iris size was adjusted each time to fulfill the filtering condition, whereas the CMOS camera allows directly record the amplitude images. In figure 4, these images are shown.

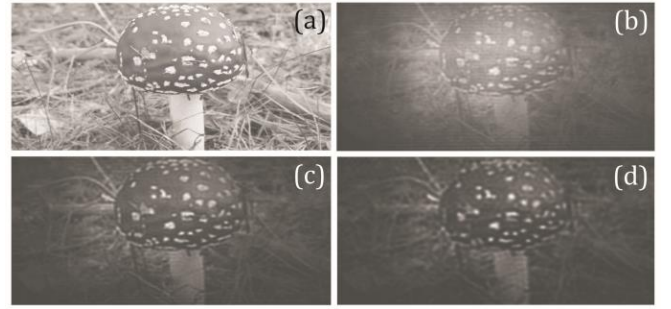

**Fig. 4.** Original amplitude pattern (a) and corresponding ones (b), (c), and (d) recorded with 1-, 4-, and 10- pixel cells, respectively.

If we compare the original image given in Fig. 4(a) with the remaining ones, it is clear that image quality is affected by the non-uniform phase response of the SLM. At the Nyquist limit (see Fig. 4(b)), the recorded image has the best resolution, but its brightness is so high that contrast becomes really poor. When pixel-cell size is increased up to 16 pixels per cell, the image contrast and sharpness are greatly improved at expense of a little decrease of resolution, please see Fig. 4(c). This can be regarded as an optimal situation, because crosstalk effects are mitigated and image resolution is still acceptable. However, if pixel-cell is increased too much, like in Fig. 4(d), problems related to the loss of resolution predominate over other any potential improve in the image quality.

Finally, we use the proposed encoding method [15] to reconstruct a non-trivial complex field under the similar pixel-cell configurations of 1, 2, and 5. In this second test, the amplitude and phase of the complex field are by given by independent images of a young boy and girl, respectively, see Figure 5(a) and (b).

The polarization-based phase shifting technique [18] is applied to measure the retrieved amplitude and phase images. To obtain the interferograms, four uniform phases with steps of π/2 radians are added to the phase element $\alpha(x,y)$. After that, we periodically eliminate some pixel-cells by multiplying $\alpha(x,y)$ by an additional binary grating $M_3(x,y)$ having double of the period of $M_1(x,y)$.

At this step, the blazed grating (equal to the previous one) is added to the generated phase element before send it to the SLM. The interferograms are formed after recombining both the light coming from the eliminated pixel-cells (reference beam), and the light modulated by the SLM at the remaining pixel-cells (object beam). The main advance of this procedure is that there is no necessity of extra optical elements (like polarizers) to obtain the interferograms. However, this is done at expense of a loss of resolution of the retrieved images because the filtering process need to be more severe. That is; the filtering condition should be accommodated to the period of $M_3(x,y)$.

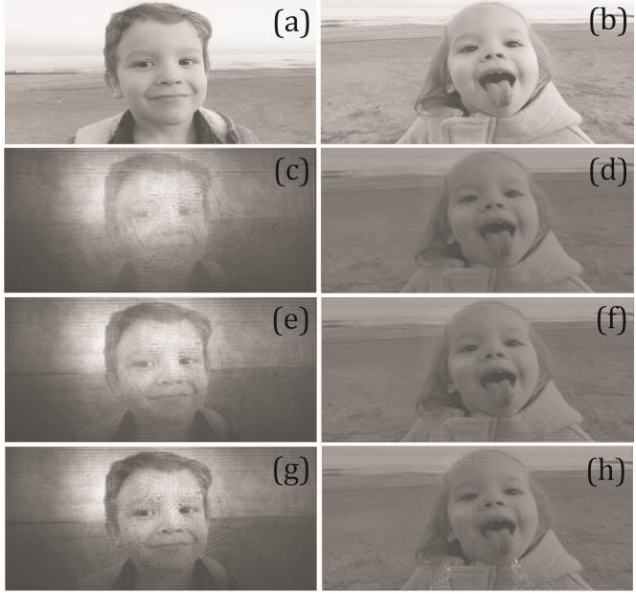

**Fig. 5.** Original amplitude (a) and phase (b) images, and corresponding amplitude (c), (e), and (g) and phase (d), (f), and (h) images recorded for sampling gratings of 1-, 2-, and 5-pixel cells, respectively.

In Fig. 5(c), (e), and (g) appear the recovered amplitude images when employing binary gratings of 1-, 2-, and 5-pixel-cells, whereas the corresponding phase images are given in Fig. 5(d), (f), and (h). From the experimental results shown in Fig. 5, one can confirm again that pixel crosstalk produces negative effects in the quality of the recorded complex field. In particular, the contrast of amplitude patterns are clearly deteriorated when decreasing the pixel-cell size. But, phase patterns seem to be poorly changed by the same reason. This last fact can be better undertook if we rewrite the phase element as:

$$\alpha(x,y) = \varphi(x,y) + \Theta(x,y) \qquad (8)$$

where,

$$\Theta(x,y) = M_1 \cos^{-1}\left[\frac{A(x,y)}{A_{max}}\right] - M_2 \cos^{-1}\left[\frac{A(x,y)}{A_{max}}\right] \qquad (9)$$

From Eq. (8) and (9) one can easily see that, in the encoding method (not in the utilized phase shifting technique), the original phase $\varphi(x,y)$ is fully encoded into the SLM. In addition, the employed optical imaging system ensures a replica of $\varphi(x,y)$ at the output plane. So, any loss of resolution of phase images should be mainly caused by the filtering process. However, the term $\Theta(x,y)$ is directly related to the encoding of amplitude information, which can only be retrieved by means of the interference among nearby pixels. So, as the interference process strongly depends on the phase values of these pixels, the amplitude images are definitively more spoiled by the pixel crosstalk effects than phase ones. These conclusions were supported by the calculus of the rmse between the original and measured patterns. It yields rmse of 23.5%, 21.6% and 19.5% for the amplitude patterns encoded with binary gratings of 1-, 2-, and 5-pixel cells, whereas for the corresponding phase patterns, the numbers were 7.9%, 8.2% and 9.1%.

**Funding.** Ministerio de Economía y Competitividad (MINECO) (FIS2016-75618-R); Generalitat Valenciana (PROMETEO/2016/079).

**Acknowledgment**. The authors are also very grateful to the SCIC of the Universitat Jaume I for the use of the femtosecond laser.